\documentclass[12pt]{iopart}

\usepackage{graphicx}
\usepackage{epsfig,rotating}
\usepackage[dvips]{color}

\def\be{\begin{equation}}
\def\ee{\end{equation}}
\def\ba{\begin{eqnarray}}
\def\ea{\end{eqnarray}}

\def\lsim{\raise0.3ex\hbox{$\;<$\kern-0.75em\raise-1.1ex\hbox{$\sim\;$}}}
\def\gsim{\raise0.3ex\hbox{$\;>$\kern-0.75em\raise-1.1ex\hbox{$\sim\;$}}}

\def\ap{\approx}

\def\eps{\varepsilon}
\def\theta{\vartheta}

\begin{document}

\title[UHECR observations and lensing in the magnetic field of the Virgo cluster]{UHECR observations and lensing in the magnetic field of the Virgo cluster}
\author{K.~Dolag$^1$, M.~Kachelrie\ss$^2$ and D.~V.~Semikoz$^{3,4}$}

\address{$^1$ Max-Planck-Institut f\"ur Astrophysik, P.O. Box 1317, D--85741 Garching, Germany}
\address{$^2$~Institutt for fysikk, NTNU, N--7491 Trondheim, Norway}
\address{$^3$~APC, 10, rue Alice Domon et Leonie Duquet, F--75205
  Paris Cedex 13, France}
\address{$^4$~INR RAS, 60th October Anniversary prospect 7a,
  117312 Moscow, Russia}

\begin{abstract}
We discuss how lensing by magnetic fields in galaxy clusters affects 
ultrahigh energy cosmic ray (UHECR) observations. As specific example,
we use Virgo together with the cluster magnetic fields obtained earlier 
in a constrained simulation of structure formation including MHD processes.
We find that, if M87 is the single source of UHECRs
from Virgo, the emitted flux is strongly anisotropic in the most 
interesting energy range, (50--100)\,EeV, and differs from the average 
value by a factor five or more for a significant fraction of observers. 
Since magnetic lensing is energy dependent, the external energy spectrum as seen
by different observers varies strongly too. These anisotropies are
averaged out in the case that all active galactic nuclei
in Virgo emit UHECRs. In both cases, the anisotropies of the emitted 
UHECR flux may introduce an important bias in the
interpretation of UHECR data like, e.g., the determination of the source
density $n_s$ and the source energy spectrum of UHECRs.
\end{abstract}


\section{Introduction}
The identification of the sources of ultrahigh energy cosmic rays
(UHECRs) is one of the most longstanding problems of
astrophysics. Only recently, evidence has been accumulating that we
are at the dawn of ``charged particle astronomy.'' A first piece of evidence
have been the observation of the Greisen-Zatsepin-Kuzmin cutoff~\cite{GZK}
in the UHECR spectrum  with $5 \sigma$ significance 
by the HiRes experiment~\cite{spectrum_HiRes} which was recently 
confirmed by the AUGER experiment with 
$6\sigma$ significance~\cite{auger_spectrum2008}.
Secondly,  anisotropies on medium scales have been found previously
combining all available data of ``old'' CR experiments~\cite{msc}.
 The data from the Pierre Auger
Observatory (AUGER) presented in Ref.~\cite{ICRC} confirmed these findings,
showing also a surplus of clustering compared to an isotropic distribution
in the broad range from 7 to 30~degrees. Such anisotropies were 
predicted as a consequence of the observed large-scale structure of
matter~\cite{old} and favor therefore together with the presence
of the Greisen-Zatsepin-Kuzmin cutoff the extragalactic origin
of UHECRs.
Comparing the expected angular auto-correlation function of different
sources with the data, the authors of Ref.~\cite{ATF} suggested
that the sources of UHECRs are active galactic nuclei (AGN) or another
strongly clustered sub-sample of galaxies.
Finally, the Auger collaboration reported  evidence for an anisotropy
of the UHECR arrival directions, observing a correlation of the
arrival directions of UHECRs with the positions of AGN or similarly 
clustered matter in the nearby Universe~\cite{auger_corr}.  
At present the significance of this correlation is only a the $3\sigma$
confidence level and is not confirmed by the data from the HiRes
experiment~\cite{HiRes}. 

Most (auto-) correlation analyses assume identical sources,
emitting isotropically UHECRs. Here we show that these assumptions
are not fulfilled in the case that UHECR sources are located 
inside regions with relatively strong magnetic fields as in the
core of galaxy clusters. As a concrete example we use the
Virgo cluster and investigate how lensing effects of magnetic fields
extending on cluster scales influence the emitted flux and energy spectrum
both  for isotropic and jet-like emission of single sources.
After a brief discussion of cluster magnetic fields in Sec.~2, 
we examine in Sec.~3 first the anisotropies in the case of a single 
isotropic source, then of jet-like emission by a single source, of
several isotropic sources and finally the resulting modifications
of the observed CR energy spectrum. We conclude in Sec.~4.

\section{Magnetic fields in galaxy clusters}

Magnetic fields have been detected in galaxy clusters by radio
observations, via the Faraday rotation signal of the magnetized
cluster atmosphere towards polarized radio sources in or behind
clusters \cite{2002ARA&A..40..319C} and from diffuse
synchrotron emission of the cluster atmosphere (see
Ref.~\cite{2008SSRv..134...93F,2004IJMPD..13.1549G} for recent
reviews). Although our understanding of their origin is still
limited, at present, numerical simulations which follow the
amplification of weak magnetic seed fields by structure formation
lead to a consistent picture of the predicted magnetic field and
its structure within galaxy clusters
\cite{1999A&A...348..351D,weak,2005ApJ...631L..21B,2008A&A...482L..13D},
well in line with the observed Faraday rotation signal. It is
worth mentioning that the predicted magnetic field structure
within galaxy clusters is found to not depend on details within
the structures of the magnetic seed fields and therefore can be
assumed to be independent on the exact mechanism which creates the
magnetic seed field at high redshift
\cite{1999A&A...348..351D,donnert2008}.

To obtain the predicted magnetic field structure for the Virgo
cluster we used the results from a constrained realization of the
local universe (see \cite{weak} and references therein). In short,
the initial conditions are similar to those used by 
Ref.~\cite{2002MNRAS.333..739M} in their study of structure formation
in the local universe. The initial density fluctuations were
constructed from the IRAS 1.2-Jy galaxy survey by smoothing the
observed galaxy density field on a scale of 7 Mpc, evolving it
linearly back in time, and then using it as a Gaussian constraint
\cite{Hoffman1991} for an otherwise random realization of the
$\Lambda$CDM cosmology. The volume constrained by the IRAS
observations covers a sphere of radius $\ap 115\,$Mpc centered on
the Milky Way and therefore encloses the prominent local structure
including the Virgo cluster. This region is sampled with high
resolution dark matter particles and is embedded in a periodic box
of around 343\,Mpc length. The region outside the constrained
volume is filled with low resolution dark matter particles,
allowing a good coverage of long range gravitational tidal forces.
Many of the most prominent clusters observed locally can therefore
be identified directly with halos in the simulation, and their
positions and masses agree well with their simulated counterparts.
For performing the MHD simulations, the original high resolution
dark matter particles where split into gas and dark matter
particles with masses of $0.69 \times 10^9\; {\rm M}_\odot$ and
$4.4 \times 10^9\; {\rm M}_\odot$, respectively. The gravitational
force resolution (i.e.\ the comoving softening length) of the
simulations was set to be $10\,{\rm kpc}$, which is comparable to
the inter-particle separation reached by the SPH particles in the
dense centers of our simulated galaxy clusters.

For this work we used the magnetic field configuration obtained
from two realizations (MHDy and MHDz) starting from
different initial seed fields (for more details see Ref.~\cite{weak})
having lower (MHDy) and higher (MHDz) initial values for the magnetic
seed field. Both simulations lead to magnetic fields within the
galaxy clusters which are statistical in agreement with the still
rare Faraday rotation measures and are roughly bracketing
the allowed range of magnetic fields in galaxy clusters.

\section{Anisotropies induced by cluster fields}

We consider first the case that only a relatively small subset of all
AGN, i.e.\ of order 1--10\%, can accelerate to the highest energies,
$E\gsim 10^{20}$\,eV. In particular, the number density of sources is then
so small that in most clusters only one UHECR source is active.
A concrete example for this situation is the acceleration of UHECRs in
hot spots of radio galaxies~\cite{RB}. For instance, there are only few
radio galaxies inside the sphere with radius 70~Mpc used in the Auger
analysis~\cite{auger_corr}. The two closest radio galaxies are Centaurus~A
and M87. The latter AGN sits in the center of the Virgo cluster.

For the calculation of CR trajectories we use magnetic fields from
the two MHD simulations of Ref.~\cite{weak}. In these simulations,
the magnetic field strength decreases fast  for increasing
distance from the cluster core, from few$\times 10^{-7}$\,G at the
core over $\sim 10^{-9}$\,G at 0.5\,Mpc to $10^{-11}$\,G 
at few Mpc. We use therefore a box of size 
5\,Mpc centered at the position of M87 and calculate the CR trajectories
under the influence of the Lorentz force until the particles leave the box.
Then we prolong the trajectory to obtain the final position on a sphere
with radius $d=18\,$Mpc equal to the distance to the Earth, 
assuming that deflections outside the box can be neglected. This
implies in particular that our results do not account for the effects
of the Galactic magnetic field.
We consider as two main cases first the isotropic emission of UHECRs 
and second jet-like emission.

Before we discuss our results, we note that large deflections
close to the source are not excluded by the Auger correlation
signal. Since the extragalactic magnetic fields found in the
simulations~\cite{weak} are strongly concentrated in the core of
the galaxy clusters on scales of size $l\sim 0.5$\,Mpc, large
deviations between the emission and exit direction from the
cluster lead only to observed deflections $\theta\sim l/d$ of
few degrees or less at the distance $d=18\,$Mpc to the Virgo cluster.
More precisely, we find that the apparent size of a point source
at the center of the Virgo cluster increases from $0.3^\circ$ at 100\,EeV
over $1.4^\circ$ at 10\,EeV to $3.6^\circ$ at 1\,EeV for an average
observer, with sizeable deviations depending on the chosen direction.

\subsection{Isotropic emission of a single source}

\begin{figure}
\begin{center}
\epsfig{file=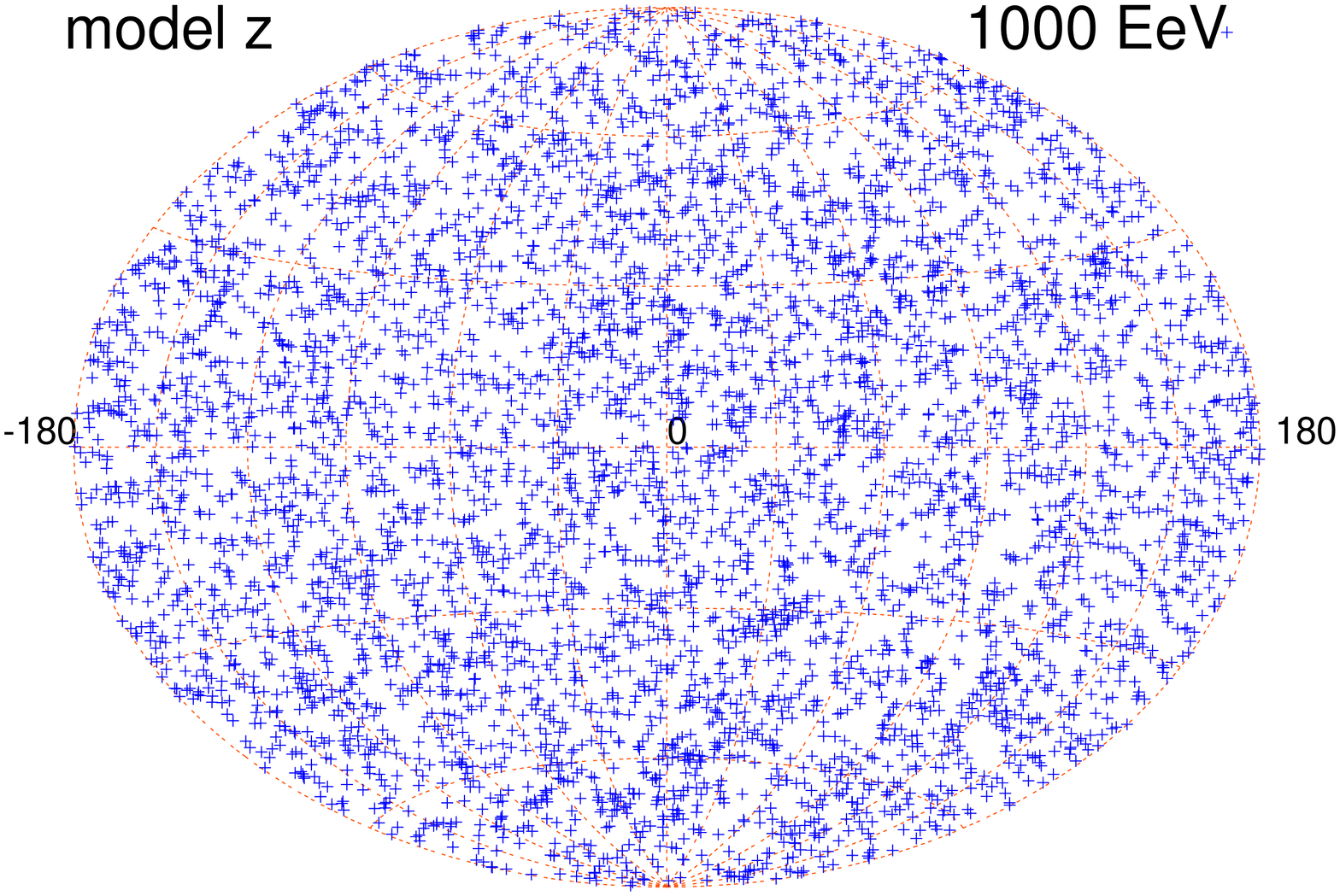,width=0.3\textwidth,angle=270}
\epsfig{file=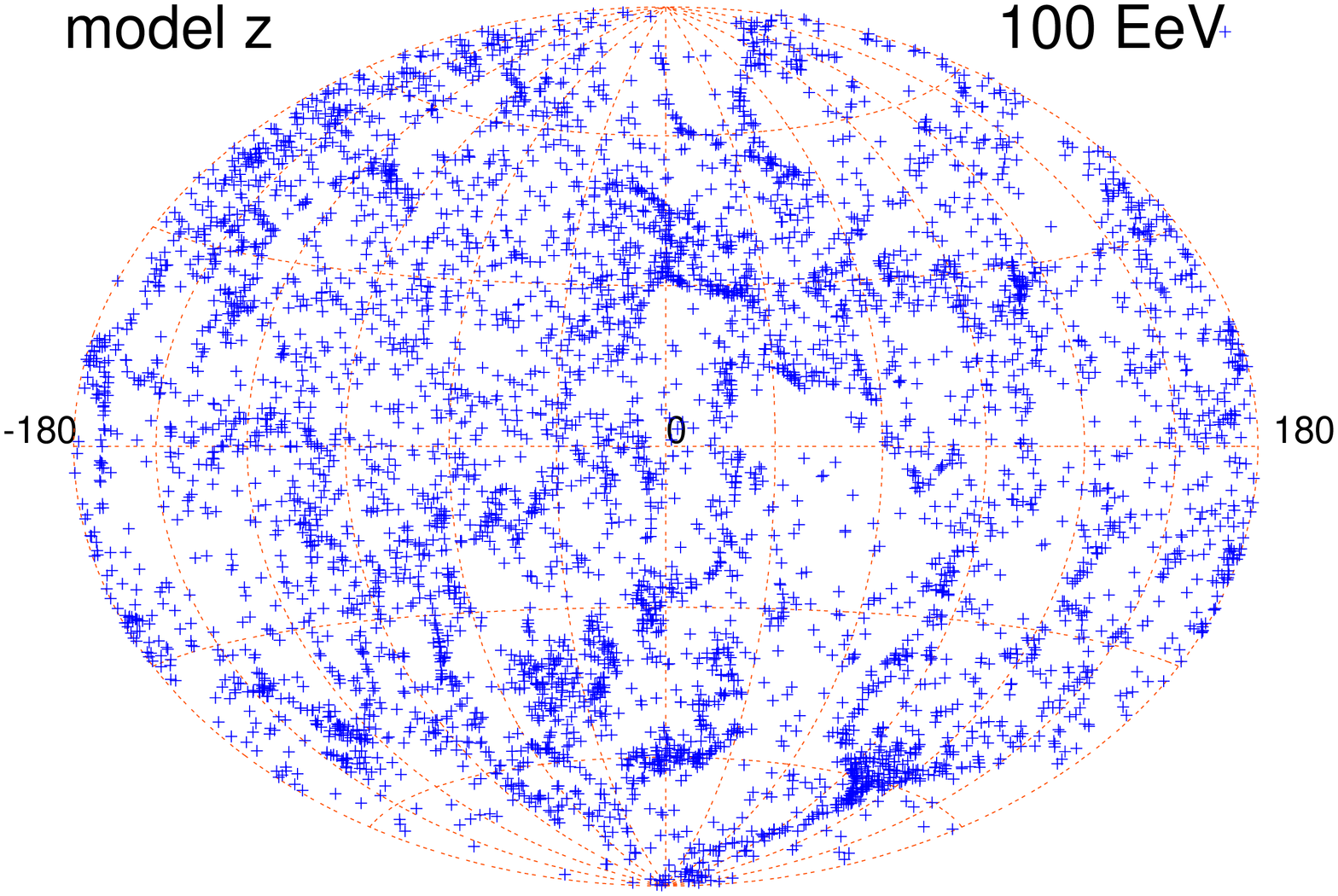,width=0.3\textwidth,angle=270}
\end{center}
\begin{center}
\epsfig{file=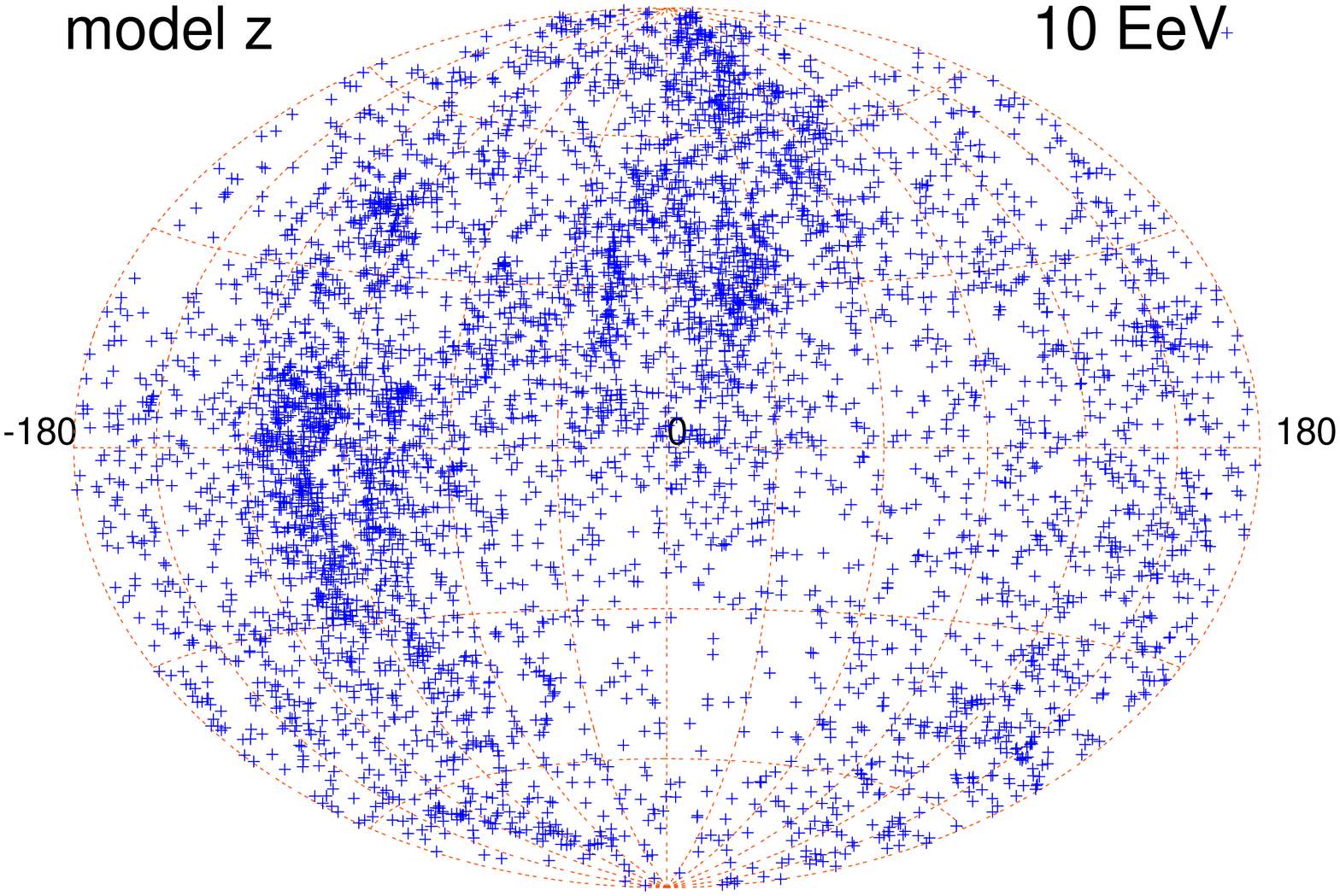,width=0.3\textwidth,angle=270}
\epsfig{file=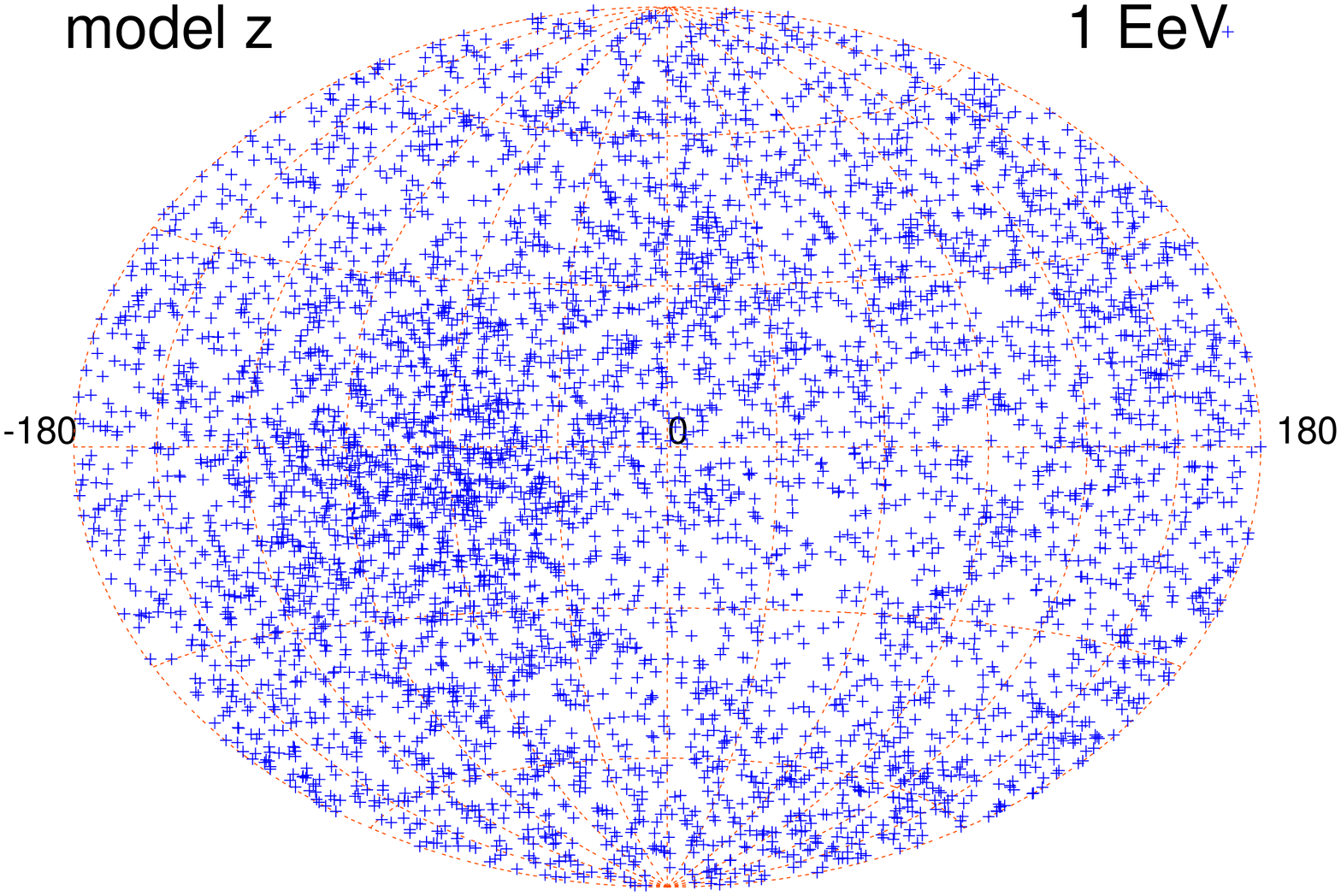,width=0.3\textwidth,angle=270}
\end{center}
\caption{Skymaps of UHECR protons emitted uniformly from M87 with
energies $E=1000$ (top, left) 100, 10, and  1\,EeV
(lower right panel).
\label{map_proton}}
\end{figure}

We inject a large number of CRs with fixed energy $E$ at the position of
M87, choosing their initial propagation direction isotropically.
Then we propagate each CR until it leaves a sphere of radius $R=2.5\,$Mpc
around M87. We neglect all energy losses during the propagation, focusing
only on the deflections due to the magnetic field. This is justified,
because the average path-length in the box is even for iron
at $E=10\,$EeV only 17\,Mpc.
Skymaps of the exit directions from this sphere are shown in
Fig.~\ref{map_proton} for protons with four different energies,
$E=1000$\,EeV (upper left panel), 100\,EeV, 10\,EeV, and 1\,EeV
(lower right panel),  using the simulation with stronger magnetic fields
(MHDz).
Large anisotropies are clearly visible by eye in the most interesting
energy range, (10--100)\,EeV.

To quantify the probability to observe a flux modified by magnetic lensing
we use the following procedure: For any random exit
directions around M87 we define
small areas of size $A\ap\pi\delta^2$ and calculate the
number $N_i$ of CRs per area and the average value $\langle N\rangle$.
Then we count the fraction $f(\eps)$ of areas where the relative number
$N_i/\langle N\rangle$ of CRs  is below or above a certain threshold value
$\eps$. The function $f(\eps)$ depends on the chosen resolution $\delta$,
i.e.\ on the size $A=\pi\delta^2$ of the areas used for averaging,
$f_\delta(\eps)$. 

A CR experiment collects during the course of a year CRs from 
Virgo within an opening angle $\theta\ap 2{\rm AU}/18{\rm Mpc}\ap 10^{-13}$ 
on a sphere centered in Virgo and radius 18\,Mpc.
We calculate therefore $f_\delta(\eps)$ for various 
values of $\delta$ and extrapolate then to $\delta\to 0$. 
This requires a large enough 
total number of cosmic rays, so that all $N_i$ are sufficiently high.
For fixed $\delta$, $\eps$ and energy $E$, both the value of
$f_\delta(\eps)$ and its uncertainty depend on the number of MC
simulations. In order to extrapolate $\delta \rightarrow 0$, we need 
to know not only $f_\delta(\eps)$, but also its fluctuations due to
the finite number of MC simulations. In order to reduce the computing time,   
we approximate the error as $df=dN/N_{tot}= f/\sqrt{(N_i)}$.
By performing many MC simulations for
several energies we checked that this approximation for $df$ is
a factor 2 or less within the exact one.

\begin{figure}
\begin{center}
\epsfig{file=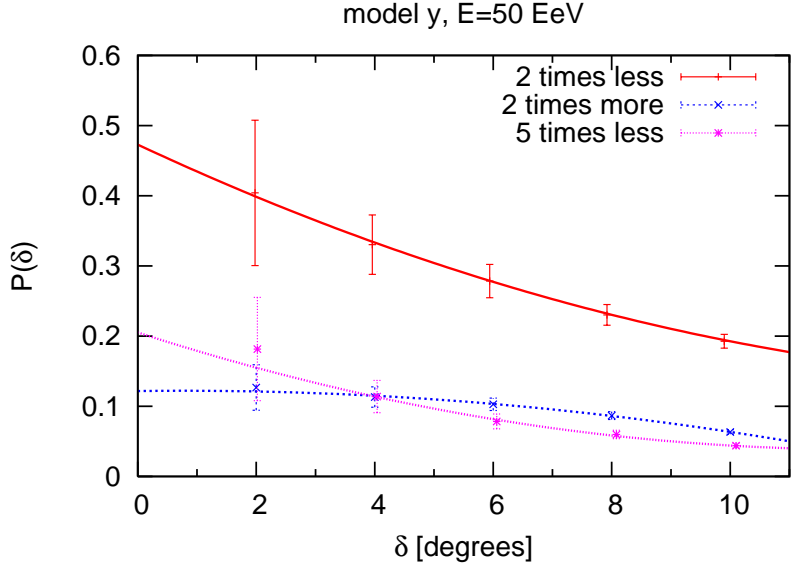,width=0.5\textwidth,angle=270}
\end{center}
\caption{ 
The fraction of regions with flux above or below a certain threshold 
value for several different viewing angles $\delta$ together with
the extrapolation curves to  $\delta\to 0$;
Red, solid line is for regions with flux twice smaller,
blue dashed line for twice larger and magenta dotted line for five times  
smaller flux than average.
\label{asy}}
\end{figure}

An example for the extrapolation  procedure  $\delta \rightarrow 0$ is 
shown in Fig.~\ref{asy}, where the
probability to observe a flux two times smaller (solid line), two times
larger (dashed line) and five times smaller (dotted line) than the average
flux is shown for model MHDy and $E=50\,$EeV for five different values of
the areas $A\ap\pi\delta^2$.  The errorbars show the error due to the
finite number of MC simulations as discussed above.
Fitting a second order polynomial $a+b\delta+c\delta^2$ to these
values, we obtain finally the asymptotic value of $f_\delta(\eps)=a$
for $\delta\to 0$ from a $\chi^2$ fit.

\begin{figure}
\begin{center}
\hskip-0.4cm
\epsfig{file=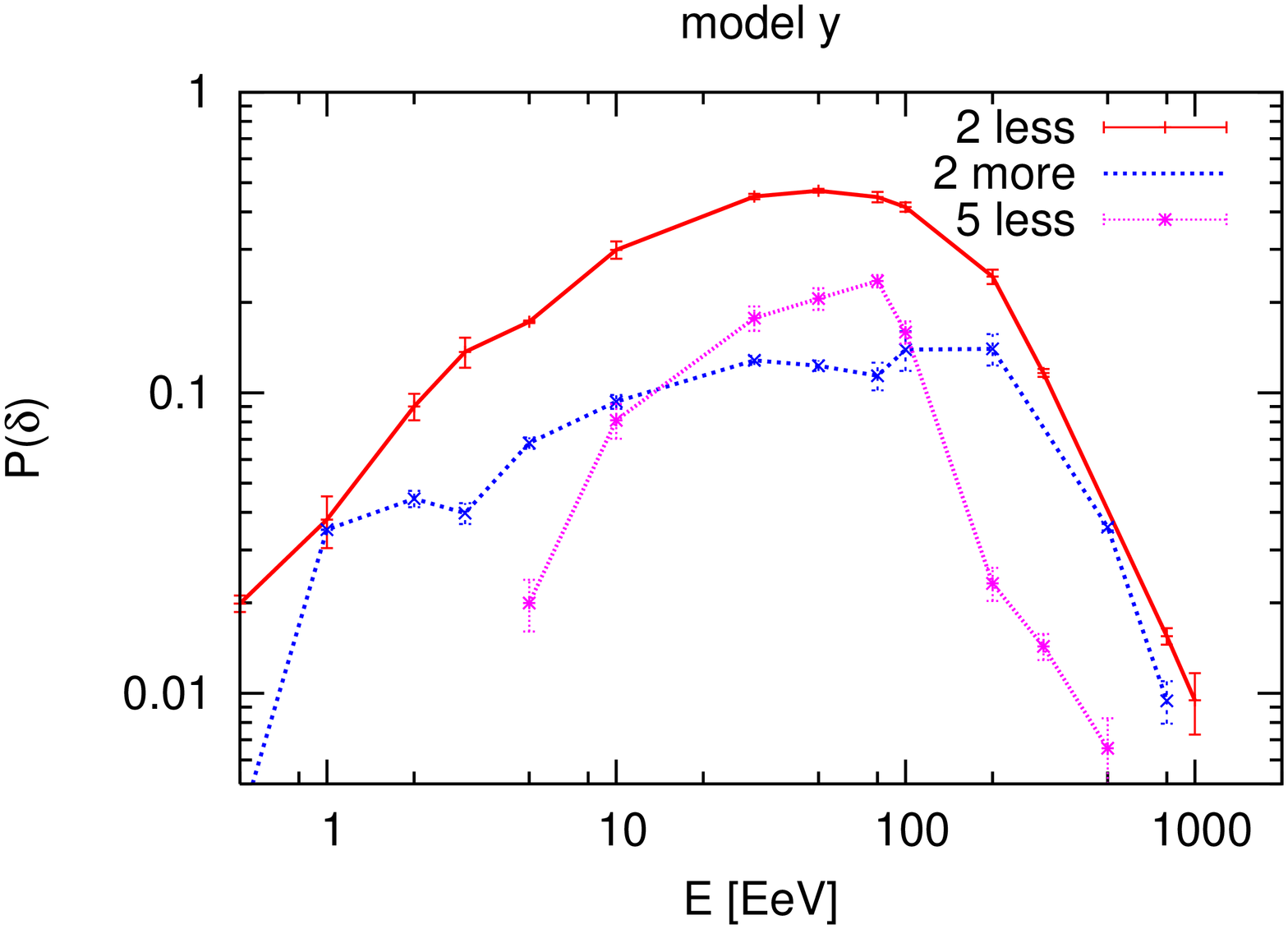,width=0.36\textwidth,angle=270}
\hskip-0.5cm
\epsfig{file=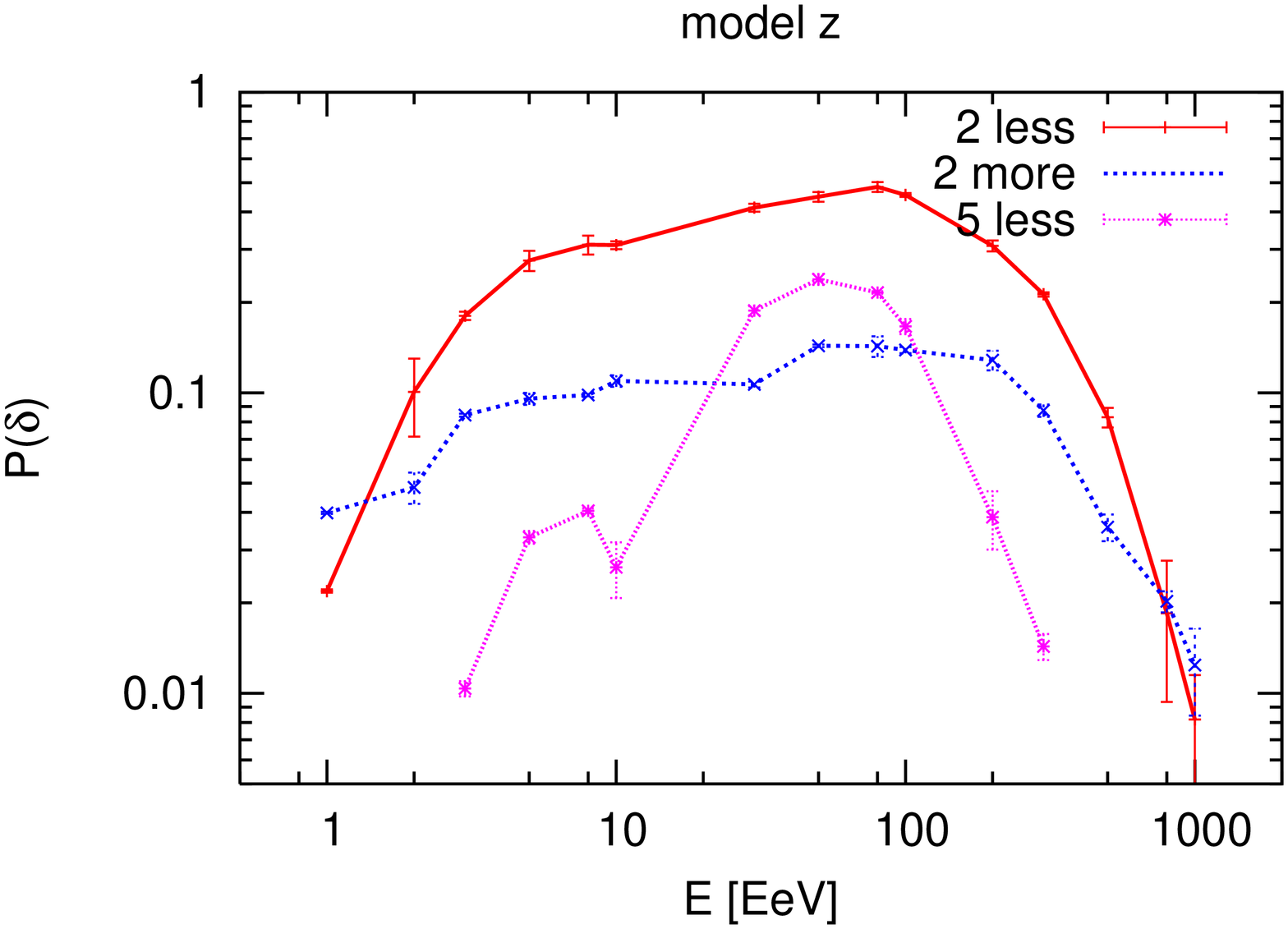,width=0.36\textwidth,angle=270}
\end{center}
\caption{The fraction of sky $f(\eps)$ exceeding the indicated level
of anisotropy, for different energies and for magnetic fields from MHDy
(left) and MHDz (right panel).
\label{frac}}
\end{figure}

The results for $f(\eps)$ obtained in this way using the
two different magnetic fields are shown in Fig.~\ref{frac}. Although
the strength of the magnetic field at the cluster center and the
deflection of individual CR trajectories differs
considerably between the two simulations, the anisotropy of the emitted
CR flux is very similar, in particular at low energies.
The reason for this result becomes clear if one
follows individual CR trajectories: The main part of the deflections is
caused by the extended field outside the cluster core, that has roughly
the same strength in both simulations.

We discuss now the results obtained for different energies.
For $E=1000$\,EeV protons propagate in the ballistic regime and deflections
are small. As a result, the sykmap of the emitted CR flux shown in the
upper left panel of Fig.~\ref{map_proton} is essentially isotropic, while
the fraction of sky with a flux modulated by a factor two is only on the
1\%~level. The situation is already qualitatively different at the for us
most interesting energy, $E=100$\,EeV. Deflections  become sizable and
regions with clearly reduced or enhanced flux are visible on the skymap, 
cf.\ the second
panel of Fig.~\ref{map_proton}. The CR flux is lower than average by a factor
two in 30\% and by a factor 5 in 5\% of the sky.

At $E=10$\,EeV protons start to propagate in the quasi-diffusive regime,
and the anisotropies are weaker. However, broad over- and underdense
regions are still visible in Fig.~\ref{map_proton}.
By contrast, at $E=1$\,EeV protons propagate in the truly diffusive regime
and most anisotropies caused by the magnetic field component are washed out:
Anisotropies on a level of a factor two occur still in 4\% of the sky,
while stronger anisotropies have such a small probability that we could
not determine them reliably ($f<10^{-3}$).

The expected anisotropies for the case of iron (charge $Q=26$)  or
other nuclei can be read off from Fig.~\ref{frac} by the
appropriate rescaling of their energy, $E\to E/Q$. Thus strong
anisotropies, e.g.\ larger than a factor five, are already
unlikely for iron nuclei with $E=100$ EeV and cluster magnetic
fields of the strength as typically predicted by the simulations
of Ref.~\cite{weak}. On the other hand, the CR flux is also for
iron nuclei modulated by a factor of two in a large fraction of
the sky.

We conclude that if M87 is the single source emitting UHECR in the Virgo
cluster, then the emission as seen from outside of the cluster is rather
anisotropic in the most interesting energy region both for iron and proton
primaries. For a significant fraction of observers the flux differs from the
average value by a factor five (protons) or two (iron), respectively.
Hence, in models assuming that only
a small fraction of all AGNs can accelerate UHECR it would be natural to
observe no UHECRs from a certain fraction of galaxy clusters, as long as the
total number of UHECR observed is as small as at present. In particular,
it would be no contradiction to the AGN source hypothesis that no UHECRs
from the Virgo cluster have been observed with the present statistics.

\subsection{Jet-like emission of a single source}

\begin{figure}
\begin{center}
\epsfig{file=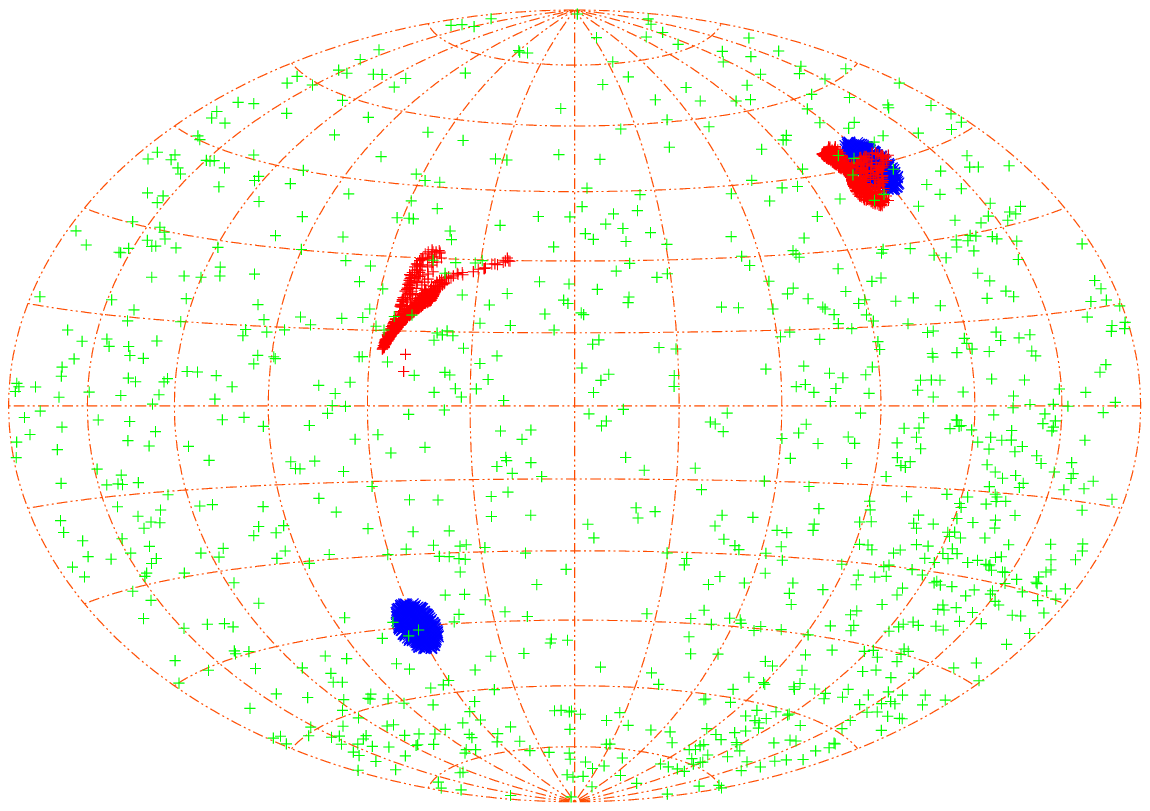,width=0.49\textwidth,angle=0}
\epsfig{file=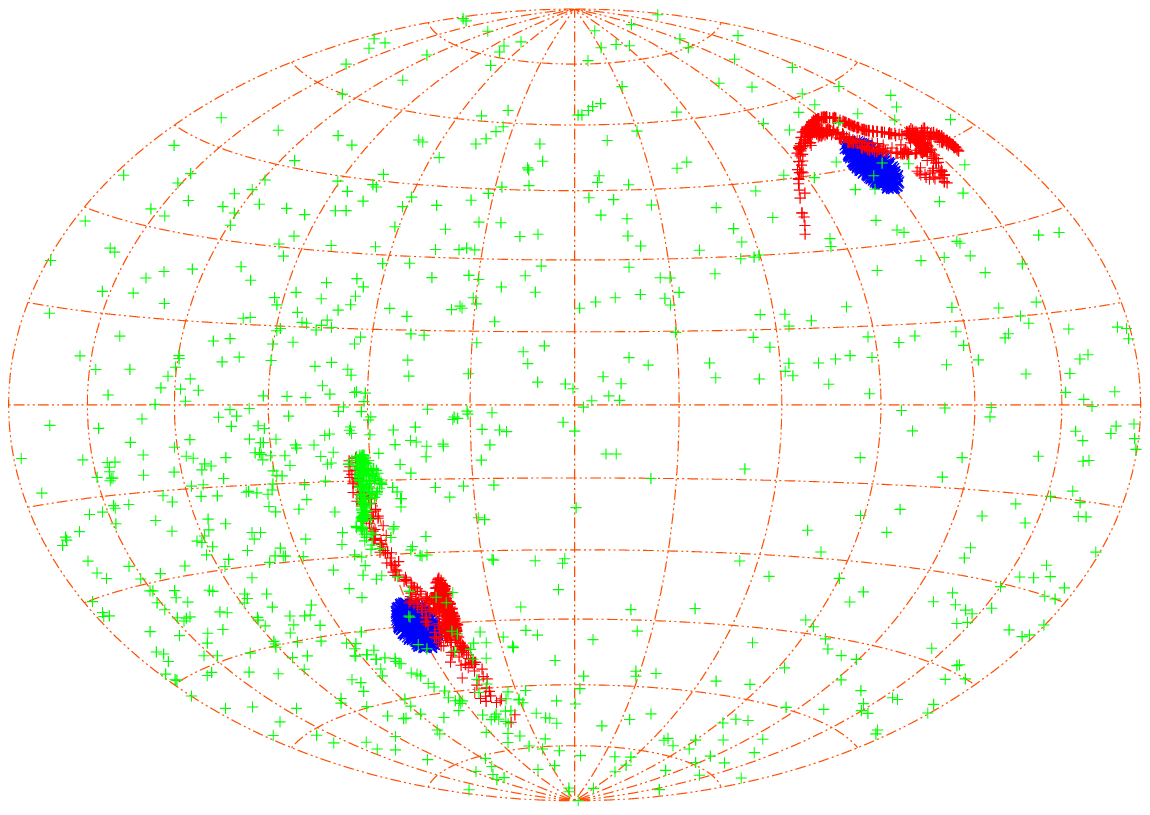,width=0.49\textwidth,angle=0}
\end{center}
\caption{Skymap of UHECRs emitted in two beams (jet-like emission)
from M87 with energy 60\,EeV, red for protons and green for iron.
The original  directions are blue, the left panel uses the magnetic
field from the MHDy and the right  from the MHDz simulation.
\label{map_jet}}
\end{figure}

We consider now the case that the acceleration of cosmic rays results in the
beamed, jet-like emission of UHECRs~\cite{jet}.
While in the previous case of isotropic emission the magnetic field
created anisotropies, its effects is now opposite, washing out
the strong initial anisotropy.
We compare again the results obtained for two different magnetic
field models from Ref.~\cite{weak}, in the left and the right
panels of Fig.~\ref{map_jet}. Each panel shows two jets with
initial opening angle $5^\circ$ in blue as well as the deflected
exit directions in red for
protons  and in green for iron, both with $E=60$\,EeV.
Both the direction and the magnitude of the deflection varies between the two
simulations and different sky regions. However, both simulations agree
that for the energy region relevant for UHECR astronomy,
$E\gsim E_{\rm GZK}\approx 50$\,EeV, the solid angle  $\Omega_{\rm eff}$
of significant UHECR
emission is increased by not more than a factor two for protons,
while the beaming of iron nuclei is destroyed nearly completely,
cf.~Fig.~\ref{map_jet}. Hence beaming
will decrease the effective number density of iron sources
only marginally, while this effect is prominent for proton sources.

Since the clustering properties of the UHECR arrival directions~\cite{ATF,ns}
and the correlation analysis~\cite{auger_corr} of the Auger collaboration,
taken at face value,
favour that the fraction of all AGN visible to us is large, the possibility
that only a small subset of all AGN like radio galaxies emits narrow
proton beams is strongly disfavoured.

\subsection{Many sources per cluster}

We assume now that all AGN emit UHECR and chose from the 12th VCV 
catalogue~\cite{VC12}
all AGN closer than 2.5 Mpc to M87 as sources. Within this distance
there are 13 additional AGN and many of them are outside the core of the
Virgo cluster
in regions with low magnetic fields. Moreover, the distance between the AGNs
is much larger than the correlation length of the field and therefore
the over- and underdense regions produced by different AGNs are un-correlated.
Consequently, the anisotropies are drastically reduced
compared to the single source case.

Assuming that all AGN have the same luminosity, we obtain a very uniform
distribution of outgoing UHECR in the case of protons primaries:
In less than 1\% of all the sky the flux deviates more than a factor two from
the average flux of UHECR protons with $E=100$\,EeV. Using iron nuclei
with the same energy increases slightly the anisotropies: The iron flux
is changed by a factor 2 in around 4\% all of areas. We stress however that
anisotropies induced by magnetic lensing for single sources lead to an 
important bias in studies of the density of UHECR sources and their energy 
spectrum, although the flux emitted by the total cluster is rather isotropic.

\subsection{Modified external CR energy spectra}

\begin{figure}
\begin{center}
\hskip-0.5cm
\epsfig{file=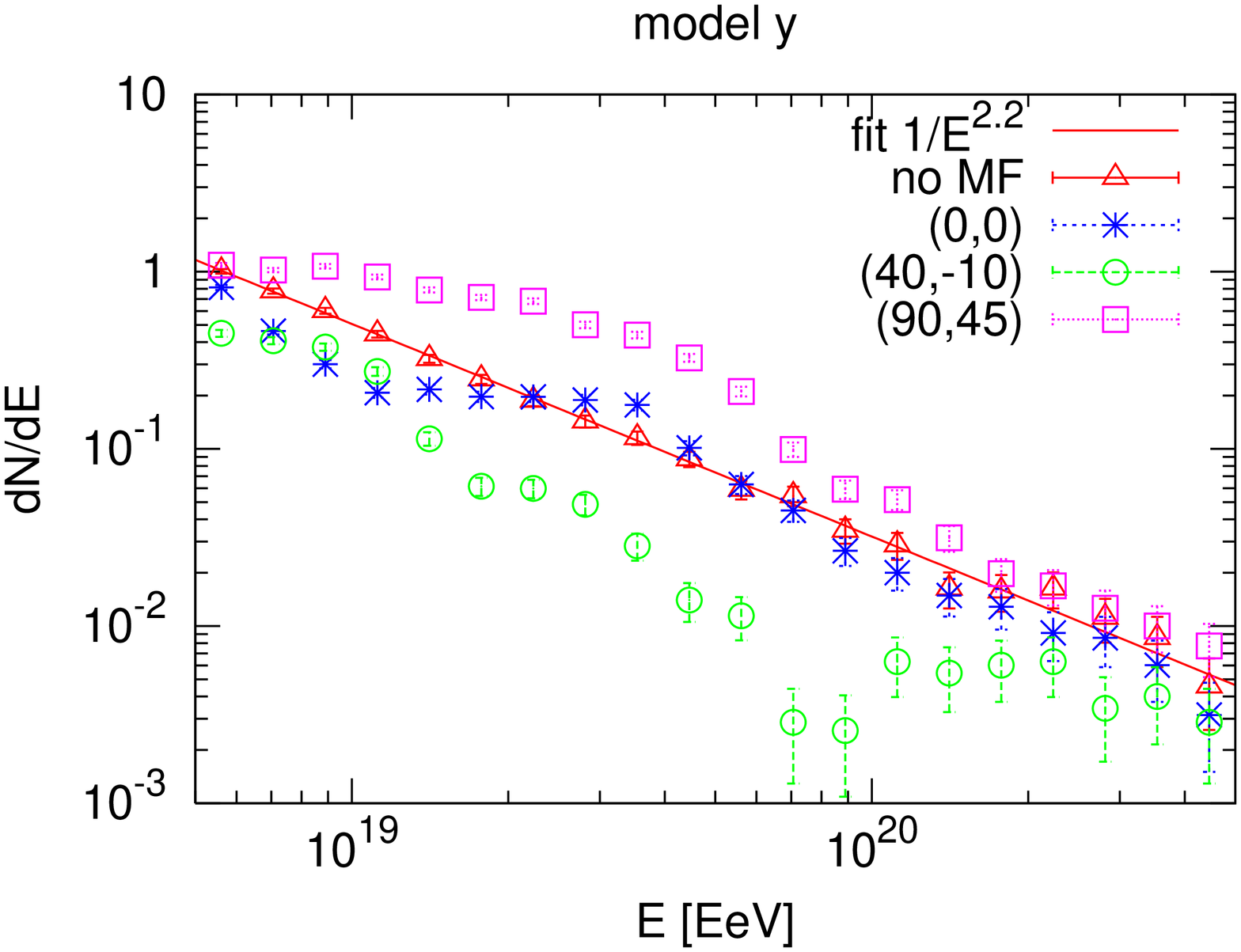,width=0.37\textwidth,angle=270}
\hskip-0.55cm
\epsfig{file=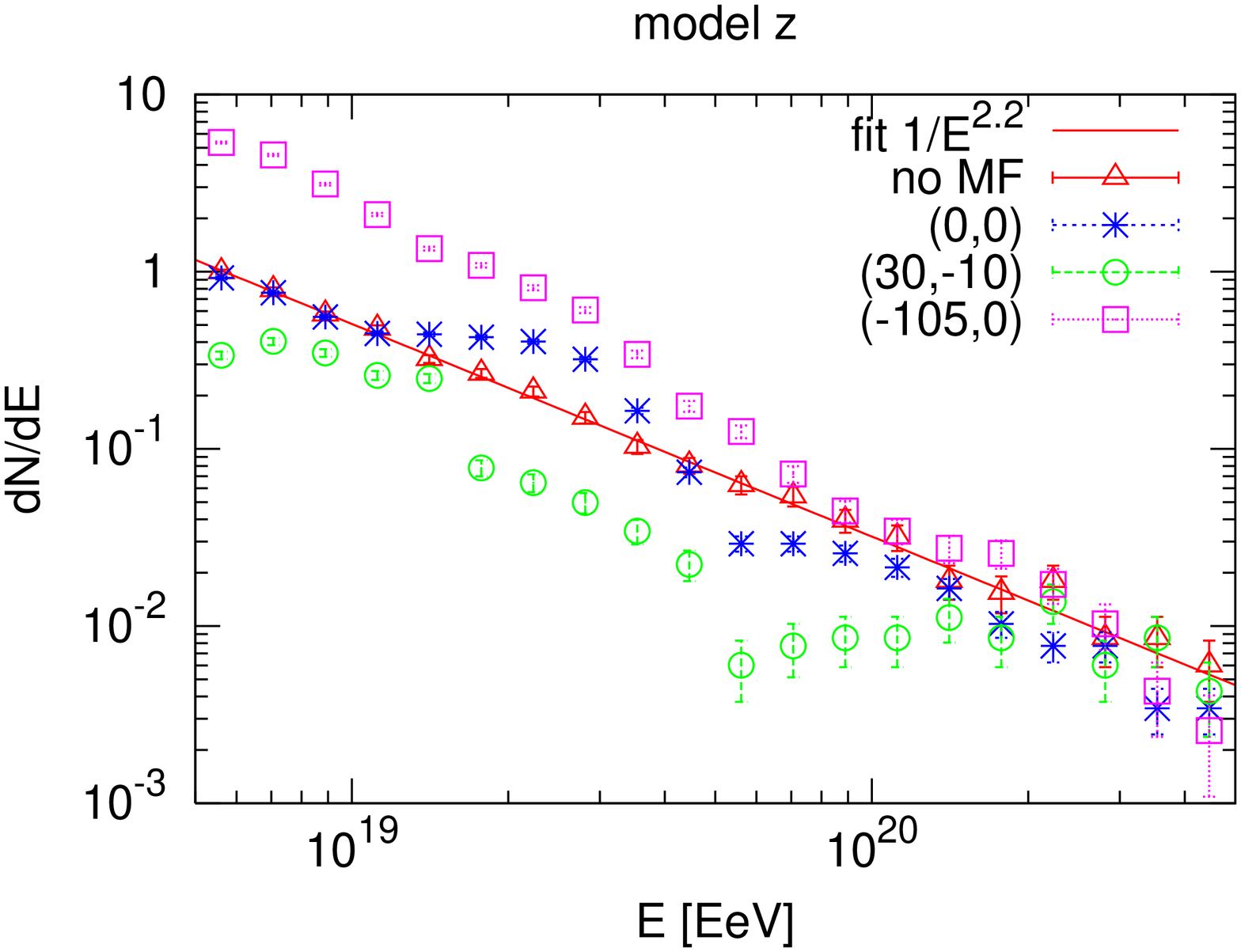,width=0.37\textwidth,angle=270}
\end{center}
\begin{center}
\epsfig{file=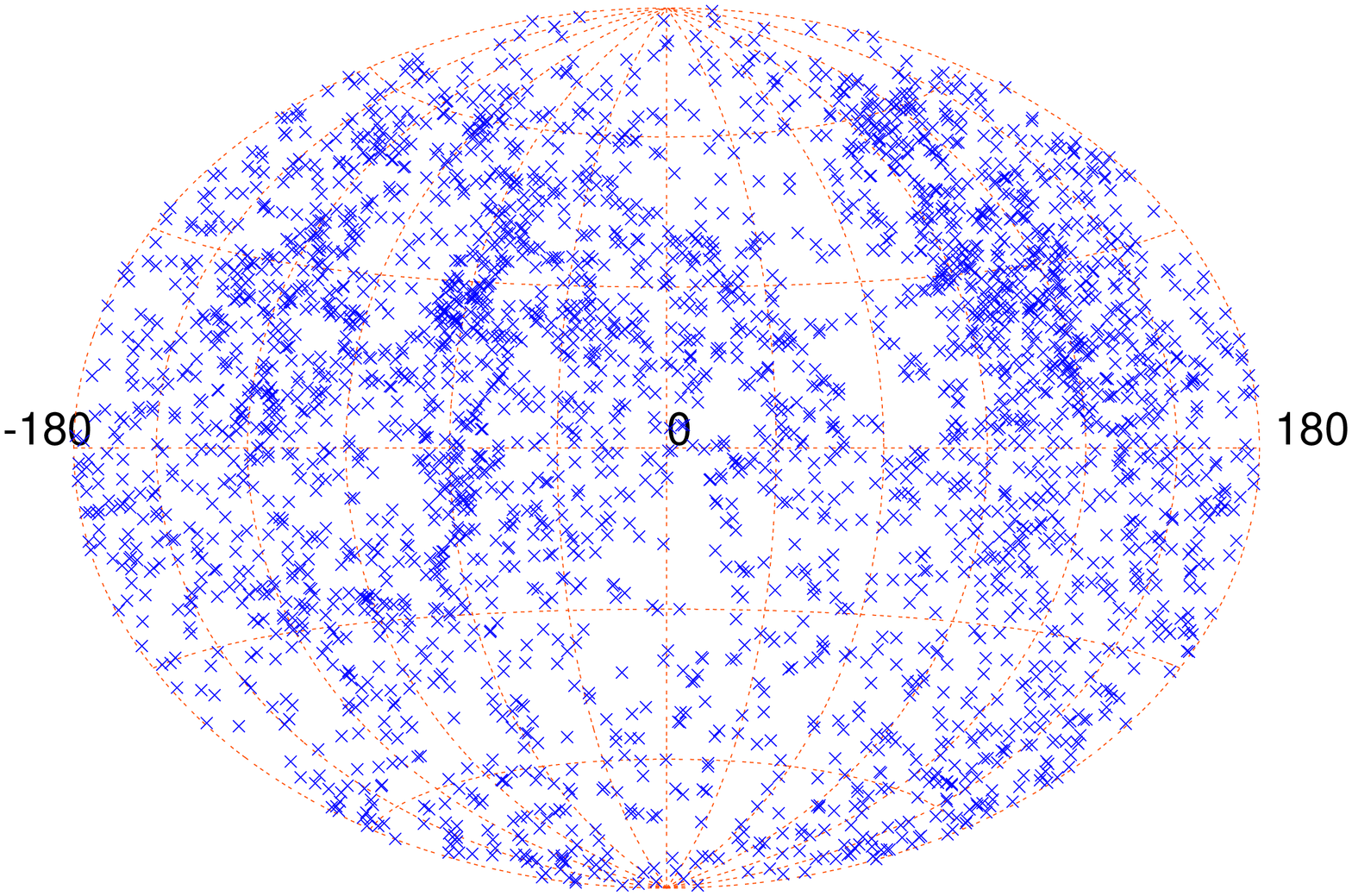,width=0.3\textwidth,angle=270}
\epsfig{file=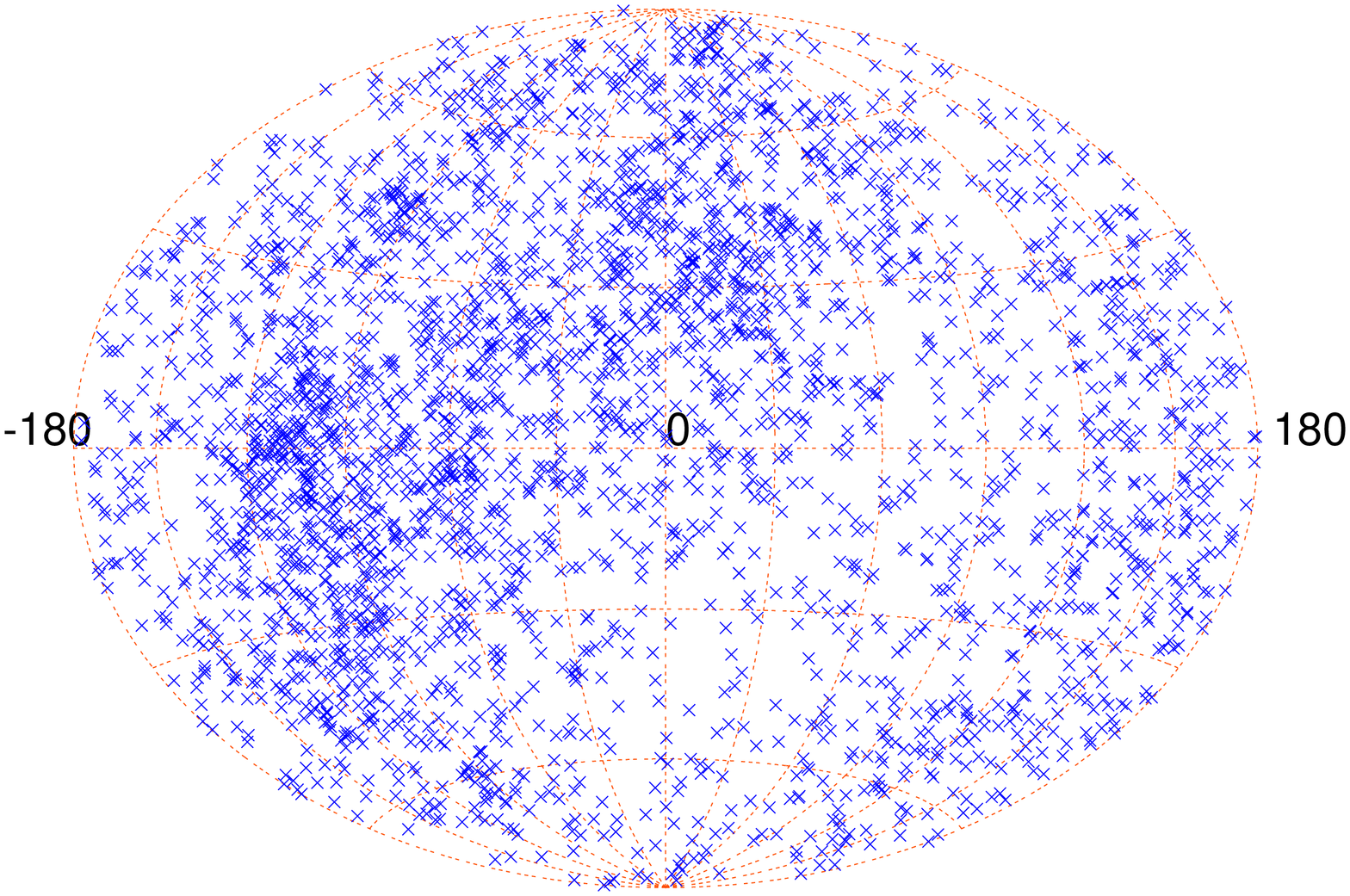,width=0.3\textwidth,angle=270}
\end{center}
\caption{The energy spectrum $dN/dE$ as function of energy (top)
and skymaps of UHECR protons emitted uniformly from M87 (bottom) for 
models MHDy (left) and MHDz (right). The energy spectrum is shown
 for $\delta=8^\circ$ regions and  $10^6$ events over the sky
 for the following cases: in absence of magnetic field
(red), spectrum in overdense region  (magenta), underdense region (green)
and ordinary region (blue). 
\label{spectrum}}
\end{figure}

Since the (de-) magnification of the CR flux by magnetic fields is
energy dependent, the anisotropies observed in the previous subsections
lead to a modulation of the initial CR spectrum for an outside 
observer.
In the Fig.~\ref{spectrum} we show the energy spectrum $dN/dE$ 
as function of energy for $10^6$ CRs and three different regions of size
$\delta=8^\circ$ in the top panels, while skymaps of 5000 UHECR
protons emitted with an  energy spectrum $dN/dE\propto E^{-2.2}$
and threshold energy $E>E_0=5$\,EeV are shown in the bottom. 
We assume first that only one source, M87, emits UHECRs.
For both models MHDy (left) and MHDz (right) we show the original
spectrum in red, in a region with average flux (blue crosses), and in two 
regions with increased (green boxes) and reduced (magenta circles) flux, 
respectively. In all cases, the spectrum is larger or smaller than average 
for several nearby bins. This mean that lensing effects are not washed out
by the finite experimental energy resolution. Also the statistical errors
shown in Fig.~\ref{spectrum} are small enough not to affect the
conclusions above. Thus, magnetic lensing does not only change the
value of the UHECR flux in a given direction, but also changes its
energy dependence:
Depending on the threshold energy and the direction considered, the exponent
of the estimated power-law $dN/dE\propto E^{-\alpha}$ may differ
strongly. Note that the shown energy spectrum will be additionally modified
both by propagation effects and by the Galactic magnetic field.
The latter will not only  deflect CRs but will lead additionally 
to a similar modulation of the CR energy spectra as the cluster field. 

\begin{figure}
\begin{center}
\hskip-0.5cm
\epsfig{file=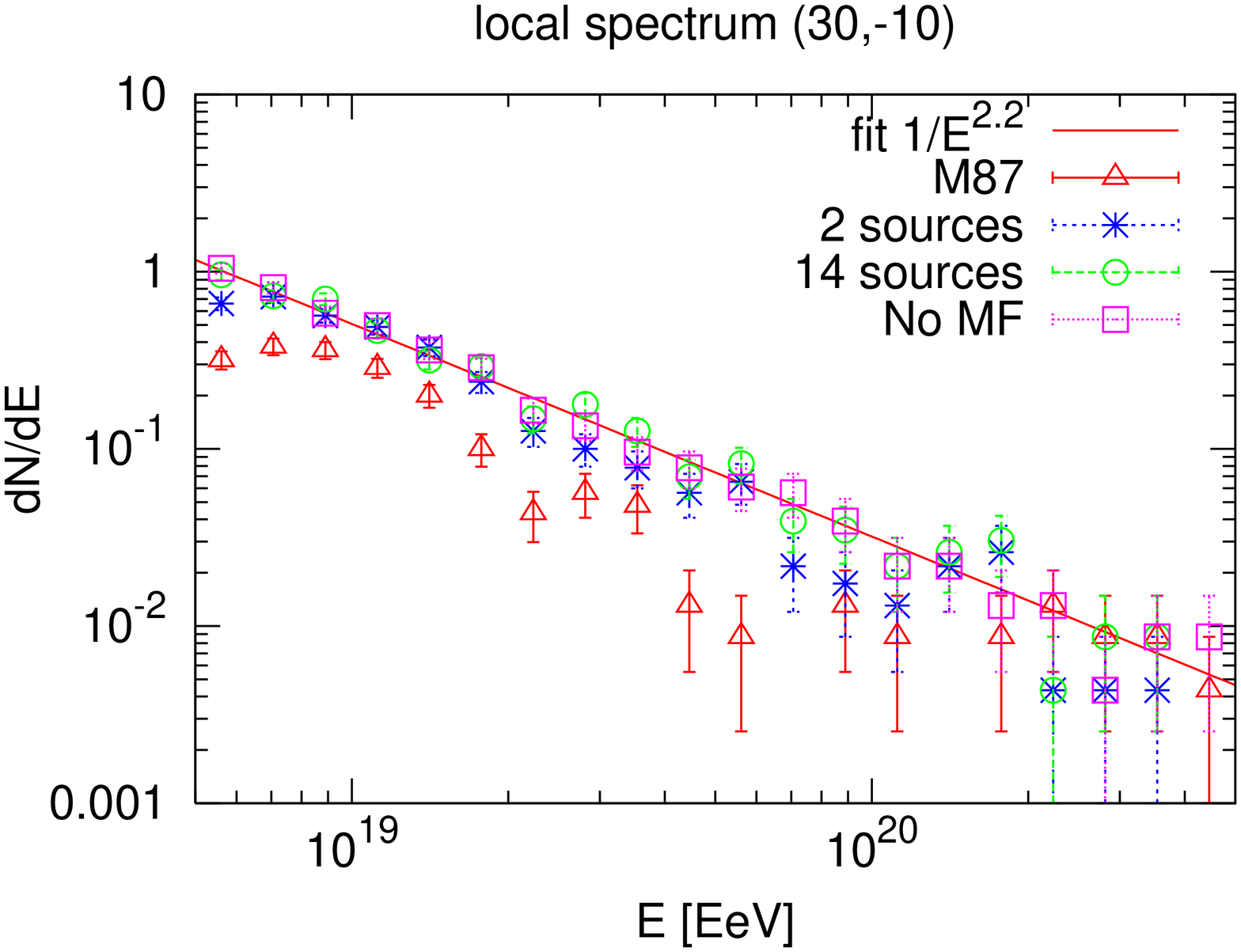,width=0.37\textwidth,angle=270}
\hskip-0.55cm
\epsfig{file=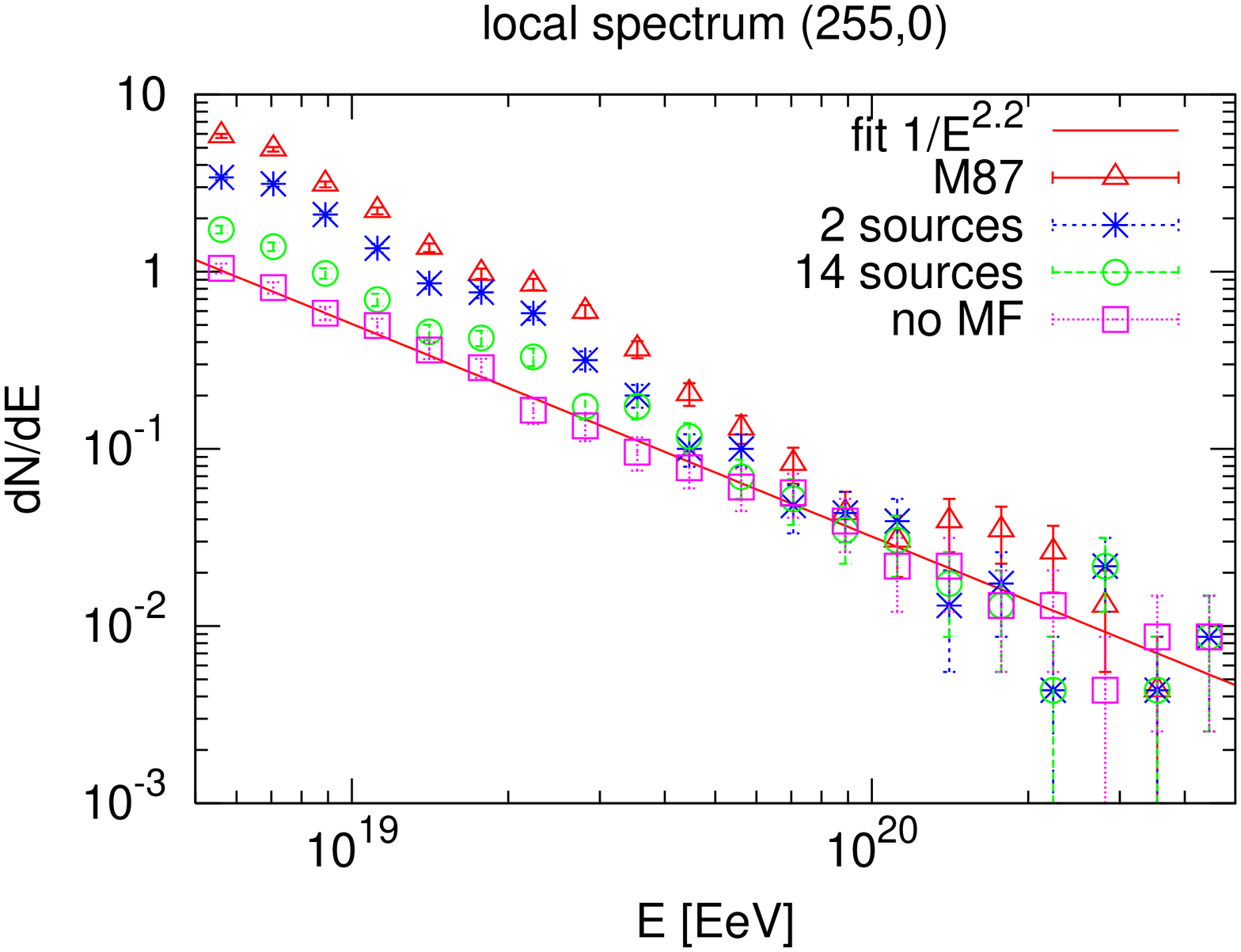,width=0.37\textwidth,angle=270}
\end{center}
\caption{
The energy spectrum $dN/dE$ as function of energy for model MHDz and
a varying number of sources; left for underdense and right for an
 overdense region. 
\label{new}}
\end{figure}

Next we investigate how fast this modification of the CR 
spectrum disappears, if the the number of sources increases. 
In Fig.~\ref{new}, we show again
the energy spectrum for two regions with reduced (left) and increased
(right) flux, varying now the number of
sources. While in the case of a spot with small anisotropy the modified 
energy spectrum resembles already for two sources the original one,
the modification of the energy spectrum in case of a large anisotropy is 
non-negligible even if all 14~AGN contribute equally to the flux. 
Finally, we remark that the case with 14 sources with equal luminosity is 
rather unrealistic, since the large difference in (bolometric) luminosity 
between M87 and the other AGN in the Virgo cluster should be reflected 
also in their UHECR luminosity.

\section{Conclusions}

The amplification of the UHECR flux by magnetic lensing was discussed
previously in the context of the Galactic regular and turbulent
field~\cite{galaxy,Kachelriess:2005qm}. For instance,
Ref.~\cite{Kachelriess:2005qm}
calculated exposure maps for three different models of the Galactic regular
magnetic field and found also significant anisotropies at $E=40\,$EeV.
The strength of the magnetic field close to the core of galaxy clusters is
smaller by a factor 30--100 than in the Milky Way. The weakness of
cluster fields is, however, over-compensated by their much larger extension
compared
to the Galactic disc. While the deflection of UHECRs in cluster fields
leads even for the nearest cluster only to maximal angular differences
between the source and the arrival direction of order $1.5^\circ$ and
is thus negligible, the resulting anisotropies of the emitted CR flux
may introduce an important bias in the interpretation of CR data.

We have found that assuming magnetic fields as in Ref.~\cite{weak}
the flux from M87 differs from the average value by a factor five
for a significant fraction of observers in the most interesting
energy region, (50--100)\,EeV. This may be the explanation
for the non-observation of UHECR from the Virgo cluster, if M87 is the
dominant UHECR source in the Virgo cluster.
More generally, anisotropies induced by magnetic fields result in a
reduction of the effective number of sources, leading to a bias in
auto-correlation studies. Vice versa, if the source type of UHECRs and thus
their number density is known, a comparison with the estimated value from
auto-correlation studies informs us about the importance of anisotropies
and/or beaming effects. Note that magnetic lensing will affect also sources 
behind the Virgo cluster.

The second important consequence of the anisotropies induced by the
cluster magnetic fields is the resulting modulation of the energy 
spectrum of UHECRs leaving the galaxy cluster compared to their generation
spectrum. A similar energy modulation is be introduced by the Galactic
magnetic field and both effects will complicate the reconstruction of
generation spectrum, if UHECRs sources are discovered.

Finally we want to stress that we obtained all our results for two
specific realizations of magnetic fields obtained within the simulations
of Ref.~\cite{weak}. Although we expect that the main features of our 
results are robust, quantitative differences will appear using different 
simulations of cluster magnetic fields: For instance, if the volume fraction 
filled by relatively large magnetic fields around clusters is as large as 
in Ref.~\cite{strong}, we expect that the same phenomena happen as found in 
this work, but extending over a larger energy range.

\section*{Acknowledgments}
 M.K.\ would like to thank the Max-Planck-Institut f\"ur Physik in 
Munich for hospitality and support.


\end{document}